\documentclass[twocolumn,showpacs,amssymb,nobibnotes,superscriptaddress,aip,10pt]{revtex4-1}
\usepackage{graphicx}
\usepackage{bm}

\begin{document}

\preprint{APS/123-QED}

\title{Highly nonclassical phonon emission statistics through two-phonon loss of van der Pol oscillator}
\author{Jiahua Li}\email{huajia\_li@126.com}
\affiliation{School of Physics, Huazhong University of Science and
Technology, Wuhan 430074, People's Republic of China}
\author{Chunling Ding}\email{clding2006@126.com}
\affiliation{Hubei Key Laboratory of Optical Information and Pattern
Recognition, Wuhan Institute of Technology, Wuhan 430205, People's
Republic of China}
\author{Ying Wu}\email{yingwu2@126.com}
\affiliation{School of Physics, Huazhong University of Science and
Technology, Wuhan 430074, People's Republic of China}
\date{\today}

\begin{abstract}
The ability to produce nonclassical wave in a system is essential
for advances in quantum communication and computation. Here, we
propose a scheme to generate highly nonclassical phonon emission
statistics---antibunched wave in a quantum van der Pol (vdP)
oscillator subject to an external driving, both single- and
two-phonon losses. It is found that phonon antibunching depends
significantly on the nonlinear two-phonon loss of the vdP
oscillator, where the degree of the antibunching increases
monotonically with the two-phonon loss and the distinguished
parameter regimes with optimal antibunching and single-phonon
emission are identified clearly. In addition, we give an in-depth
insight into strong antibunching in the emitted phonon statistics by
analytical calculations using a three-oscillator-level model, which
agree well with the full numerical simulations employing both a
master-equation approach and a Schr\"{o}dinger-equation approach at
weak driving. In turn, the fluorescence phonon emission spectra of
the vdP oscillator, given by the power spectral density, are also
evaluated. We further show that high phonon emission amplitudes,
simultaneously accompanied by strong phonon antibunching, are
attainable in the vdP system, which are beneficial to the
correlation measurement in practical experiments. Our approach only
requires a single vdP oscillator, without the need for reconfiguring
the two coupled nonlinear resonators or the complex nanophononic
structures as compared to the previous blockade schemes. The present
scheme could inspire methods to achieve antibunching in other
systems.
\end{abstract}

\pacs{42.60.Da, 42.50.Ct, 05.45.Xt, 42.65.-k}

\maketitle

\section{Introduction}\label{I}
In analogy to the photon antibunching
\cite{I-1,I-2,I-3,I-4,I-5,I-6,I-7,I-8,I-9,I-10,I-11,I-12}, the
phonon antibunching refers to the phenomenon that the coupling of a
single phonon to the system hinders the coupling of the subsequent
phonons, which is a pure nonclassical effect owing to less than
unity value of the normalized second-order correlation function of
the phonon field. The phonon antibunching is important in uncovering
the quantum behaviors of the mechanical vibration and in realizing a
single-phonon source and other single-phonon quantum devices. Here,
on the one hand, the phonons are produced by a mechanical vibration
(a motional degree of freedom) and are the outcome of the vibration
quantization \cite{I-13a,I-13b}. Generally, the mechanical vibration
frequencies are typically on the order of MHz and GHz, which is
quite different from the ones of the photons. On the other hand, the
phonons, like the photons, belong to the bosons and belong to the
category of quantum harmonic oscillators. Moreover, their mechanical
operators have the same correspondence. As a result, the signatures
of the phonon antibunching (also called as phonon blockade) can be
detected through the phonon-statistics measurements, such as the
above-mentioned second-order correlation function at zero time delay
\cite{I-14,I-15}. The realization of the phonon antibunching
generally needs a large quantum mechanical nonlinearity like the
photon antibunching. In recent years, it has been shown that this
mechanical nonlinearity can be achieved in hybrid systems with the
mechanical oscillator coupled to a single qubit
\cite{fa1,fa2,fa3,fa4} and in optomechanical systems with quadratic
or three-mode coupling \cite{fb1,fb2}. Single-photon-induced phonon
blockade has been investigated theoretically in a hybrid
spin-optomechanical system \cite{I-16}.

On the other hand, the mechanical van der Pol (vdP) oscillator is
very extensive in natural and engineered systems. Since the
prototypical model of the mechanical vdP oscillator was first
proposed by Balthazar van der Pol \cite{fc1} in 1926, it has
attracted considerable attention in the classical level. In the past
few years, this vdP model has been successfully extended from a
classical regime to a full quantum regime, i.e., the so-called
quantum vdP oscillator, and also has drawn a great deal of research
interests in the quantum level. The particles in the vdP field are
usually called the phonons based on the literatures
\cite{VI-1,VI-2,VI-3,VI-4,VI-5,VI-6,VI-7,V-8,V-9,I-79,I-80,I-81,I-82}
and the system is also called a phonon vdP oscillator. We notice
that the previous works shed light on the quantum synchronization,
limit-cycle, frequency entrainment, and diverging and negative
susceptibilities of a quantum vdP oscillator
\cite{VI-1,VI-2,VI-3,VI-4,VI-5,VI-6,VI-7,V-8,V-9} or the
entanglement tongue and quantum synchronization of two coupled
quantum vdP oscillators \cite{I-79,I-80,I-81}, differing from what
are found in the classical regime. The multiple resonances of the
mean phonon number are demonstrate in two coupled quantum anharmonic
vdP oscillators and the genuine quantum effects are also expected in
the amplitude death phenomenon \cite{I-82}. To date, there seems to
exist no works in the literature dealing with the above-mentioned
nonclassical phonon antibunching in a quantum vdP oscillator. Is
there strong antibunching effect? How does quantum noise (the
dissipative coupling) effect the antibunching?

Motivated by these facts and to answer these issues above, in the
present work we theoretically investigate the characteristics of
antibunched phonons generated in a driven quantum vdP oscillator
with both single- and two-phonon losses. By means of the
second-order correlation function $g^{(2)}(0)$ providing information
on phonon bunching and antibunching, we show that the degree of
phonon antibunching is governed sensitively by the nonlinear
two-phonon loss of the vdP oscillator. Increasing the two-phonon
loss increases the degree of the phonon antibunching with
experimentally achievable parameters. This occurs because a large
two-phonon loss decouples the third level of an oscillator leaving
an effective driven two-level system. We identify clearly
distinguished parameter regimes with strong antibunching and
single-phonon emission. On the other hand, we present analytical
solutions of the second-order correlation function based on a
three-oscillator-level approximation, which match the full numerical
simulations based on a quantum master equation and a Schr\"{o}dinger
equation at weak driving. Additionally, we analyze the fluorescence
phonon emission spectra of the vdP oscillator by properly varying
the two-phonon loss. It is found that high phonon emission
amplitudes, accompanied by strong phonon antibunching (single-phonon
emission), can be achieved efficiently. This result is useful for
the correlation measurement in practical experiments.

Compared with the previous phonon antibunching or blockade schemes
\cite{fa1,fa2,fa3,fa4,fb1,fb2}, our approach is based entirely on a
single vdP oscillator, eliminating the need for coupled nonlinear
resonators or complex nanophononic structures. On the other hand,
our achievable antibunching effect, $g^{(2)}(0)<1$, can occur,
without requiring fine trade-off and tuning between the loss rate
and the nonlinearity. In particular, the degree of phonon
antibunching can be kept high even for the case that the nonlinear
two-phonon damping is less than the linear single-phonon damping.
Finally, we briefly discuss specific implementations based
respectively on an ion trap
\cite{VI-1,VI-2,VI-3,VI-4,VI-5,VI-6,VI-7} and an optomechanical
membrane \cite{VI-8, VI-9, VI-10}. Our study contributes insights in
understanding phonon-phonon correlation in the quantum vdP field.

Finally, it is worth pointing out that the same model in the
presence of both the photon-photon (or phonon-phonon) interaction
and the two-photon (or two-phonon) driving has been proposed and
explored by means of the complex P-representation formalism by
Bartolo \emph{et. al} \cite{r1}. Many interesting features,
including controllable Wigner-function multimodality, dissipative
phase transitions, quantum trajectories, and Schr\"{o}dinger cat
states, have been elaborated in the Bartolo's model or approximate
model \cite{r1,r2,r3}. But, these are very different scenarios to
ours as we are interested in quantum statistics of the phonon
emitted in the vdP oscillator in the absence of the phonon-phonon
interaction and the two-phonon driving, namely, such a regime is not
addressed explicitly in the works mentioned before. From an
experimental point of view, this is beneficial because there is no
need for invoking the additional Kerr nonlinearity and parametrical
driving source, which makes the implementation of this proposal more
experimental friendly.

This paper is organized as follows. In Sec.\,\ref{II}, we describe
the quantum model of a driven vdP oscillator under consideration and
yield the Lindblad master equation which governs the dissipative
dynamics of the vdP oscillator. In Sec.\,\ref{III}, we introduce the
second-order intensity correlation function which provides
information on the degree of phonon bunching and antibunching.
Subsequently, in Sec.\,\ref{IV}, we provide analytical insights into
the second-order correlation function based on a
three-oscillator-level model (Sec.\,\ref{IVA}) via a master equation
approach and further find approximate solutions of the second-order
correlation function (Sec.\,\ref{IVB}). Again, starting from a
Schr\"{o}dinger equation approach, an analytical discussion for the
second-order correlation function of the system is presented
(Sec.\,\ref{IVC}). This three-oscillator-level model is a
justifiable approximation for low driving strengths. In
Sec.\,\ref{V}, we explore in details phonon correlation
characteristics without and with time delay, as well as emission
spectrum of the vdP oscillator by adjusting the typical parameters,
especially dissipative engineering the two-phonon loss. In
Sec.\,\ref{VI}, we elaborate on an ion trap and optomechanical
implementations of the quantum vdP oscillator. Finally, we summarize
our results in Sec.\,\ref{VII}.

\section{System of a quantum vdP oscillator: Hamiltonian and master equation treatment}\label{II}
We first introduce our model of the system. We consider a quantum
vdP oscillator of interest, subject to an external continuous-wave
driving of strength $E_{d}$ and frequency $\omega_{d}$, i.e.,
$S_{dri}(t)= E_{d}e^{-i\omega_{d}t}$. Our oscillator model involves
two dissipation processes, one being a linear single-phonon loss of
rate $\kappa_{1}$ and the other being a nonlinear two-phonon loss of
rate $\kappa_{2}$, as in
Refs.\,\cite{VI-1,VI-2,VI-3,VI-4,VI-5,VI-6,VI-7}. Nevertheless, our
model makes a slight adjustment of the usual quantum vdP oscillator,
where the single-phonon gain is replaced with the single-phonon loss
since the presence of the single-phonon loss is preferred to obtain
strong phonon antibunching, and, on the other hand, it is much
easier to achieve experimentally. Such dissipation processes can be
engineered in current experimental platforms by coupling the
oscillator to a suitable environment, as will be discussed later.
The Hamiltonian and the Lindblad master equation for this model, in
the frame rotating at the frequency $\omega_{d}$ of the external
driving, read (setting $\hbar=1$ throughout this work)
\begin{eqnarray}
\hat{H}_{osc}&=&\hat{H}_{0}+\hat{H}_{dri}=\Delta\hat{a}^{\dag}\hat{a}+iE_{d}(\hat{a}-\hat{a}^{\dag}), \label{1}\\
\frac{d\hat{\rho}}{dt}&=&-i[\hat{H}_{osc}, \hat{\rho}]
+\frac{1}{2}\kappa_{1}\mathcal{D}(\hat{a})\hat{\rho}+\frac{1}{2}\kappa_{2}\mathcal{D}(\hat{a}^{2})\hat{\rho},
\label{2}
\end{eqnarray}
where $\hat{a}$ ($\hat{a}^{\dag}$) is the oscillator annihilation
(creation) operator satisfying bosonic commutation relation
$[\hat{a}, \hat{a}^{\dag}]=1$, $\Delta=\omega_{a}-\omega_{d}$ is the
detuning between the vdP oscillator frequency $\omega_{a}$ and the
external driving frequency $\omega_{d}$, and $\hat{\rho}$ is the
density matrix operator, respectively. Applying the unitary
transformation $\hat{U}=\exp(i\omega_{d}\hat{a}^{\dag}\hat{a}t)$
leads to Eq.\,(\ref{1}) above:
$\hat{H}_{osc}=\hat{U}\hat{H}_{old}\hat{U}^{\dag}-i\hat{U}\partial\hat{U}^{\dag}/\partial
t$, where
$\hat{H}_{old}=\omega_{a}\hat{a}^{\dag}\hat{a}+iE_{d}(e^{i\omega_{d}t}\hat{a}-e^{-i\omega_{d}t}\hat{a}^{\dag})$,
under the helps of
$\hat{U}\hat{a}\hat{U}^{\dag}=\hat{a}e^{-i\omega_{d}t}$,
$\hat{U}\hat{a}^{\dag}\hat{U}^{\dag}=\hat{a}^{\dag}e^{i\omega_{d}t}$,
and $\hat{U}\partial\hat{U}^{\dag}/\partial
t=-i\omega_{d}\hat{a}^{\dag}\hat{a}$. In the driving term
$\hat{H}_{dri}=iE_{d}(\hat{a}-\hat{a}^{\dag})$ in Eq.\,(\ref{1}),
the rotating-wave approximation has been applied; without the loss
of generality, we set $E_{d}$ to be real.
$\hat{H}_{0}=\Delta\hat{a}^{\dag}\hat{a}$ is the unperturbed
Hamiltonian of the oscillator relative to $\omega_{d}$. Above, in
Eq.\,(\ref{2}) we have introduced the Lindblad operator
$\mathcal{D}(\hat{\mathcal{O}})\hat{\rho}=2\hat{\mathcal{O}}\hat{\rho}\hat{\mathcal{O}}^{\dag}
-\hat{\rho}\hat{\mathcal{O}}^{\dag}\hat{\mathcal{O}}
-\hat{\mathcal{O}}^{\dag}\hat{\mathcal{O}}\hat{\rho}$
to describe the nonunitary dynamics with $\hat{\mathcal{O}}$ being
the relevant operators $\hat{a}$ and $\hat{a}^{2}$. This master
equation can be used to derive the steady state for the following
second-order correlation function.

\section{Second-order correlation function providing information on phonon (anti)bunching}\label{III}
The appeal of the quantum vdP oscillator is that owing to its simple
form, it can serve as a prototypical model for studying its
nonclassical properties in the quantum limit. Here we concentrate on
the phonon statistics of the quantum vdP oscillator. To this end,
our central task in this work is to calculate the normalized
second-order correlation function $g^{(2)}(\tau)$, which is defined,
for the phonon vdP oscillator in the steady state, as
\cite{V-5,V-6,V-7,V-10}
\begin{eqnarray}\label{3}
g^{(2)}(\tau)&=&\frac{\mbox{Tr}[\hat{\rho}\hat{a}^{\dag}(0)\hat{a}^{\dag}(\tau)\hat{a}(\tau)\hat{a}(0)]}
{\{\mbox{Tr}[\hat{\rho}\hat{a}^{\dag}(0)\hat{a}(0)]\}^{2}}\nonumber\\
&=&\frac{\langle\hat{a}^{\dag}(0)\hat{a}^{\dag}(\tau)\hat{a}(\tau)\hat{a}(0)\rangle}{\langle\hat{a}^{\dag}(0)\hat{a}(0)\rangle^{2}},
\end{eqnarray}
where
$\langle\hat{\mathcal{O}}\rangle=\mbox{Tr}(\hat{\rho}\hat{\mathcal{O}})$
denotes the expectation values taken with respect to the
steady-state density matrix of the full Lindblad master equation
(\ref{2}), and $\tau$ is the time between the arrival of the first
and second phonons, i.e., time delay. $g^{(2)}(\tau)$ can be
understood as the conditional probability of detecting a phonon at
time $\tau$ provided that one phonon was detected earlier at time
$\tau=0$. It can be proved that the second-order correlation
function $g^{(2)}(\tau)$ satisfies the properties of, on the one
hand, time symmetry $g^{(2)}(-\tau)=g^{(2)}(\tau)$, and, on the
other hand, non-negativity $g^{(2)}(\tau)\geq0$.

If $\tau=0$, we refer to $g^{(2)}(0)$ as a zero-time-delay
second-order correlation function. For the case of zero time delay,
it is known that the value of $g^{(2)}(0)<1$ [$g^{(2)}(0)>1$]
corresponds to sub-Poissonian (super-Poissonian) statistics of the
vdP oscillator field, which is a nonclassical antibunching
(classical bunching) effect. Alternatively, $g^{(2)}(0)=1$
represents a coherent light source.

One can numerically calculate the steady-state solution of the
master equation Eq.\,(\ref{2}) with the left-hand side set to zero,
from which the steady-state second-order correlation function
$g^{(2)}(\tau)$ is obtained. For this purpose, we truncate the vdP
field's Hilbert space sufficiently large, for example, at phonon
numbers as large as $100$ for the vdP oscillator mode to ensure full
convergence. Finally, all calculations are performed in the frame
rotating with the driving frequency $\omega_{d}$.

\section{Two-type analytical insights into the second-order correlation function by a three-oscillator-level approximation} \label{IV}
\subsection{Three-oscillator-level approximation via a master equation approach} \label{IVA}
Before proceeding, we develop a three-oscillator-level model that
provides further physical insight into the numerical results. For
this purpose, in the Hilbert space of Fock states $\{|n\rangle\}$
with $n=0, 1, 2, \ldots$ being the number of phonons in the
oscillator mode, the Lindblad master equation (\ref{2}) can be
written as
\begin{eqnarray}\label{4}
\frac{d\rho_{n,m}}{dt}&=&-i\Delta\left(n-m\right)\rho_{n,m} \nonumber\\
&+&E_{d}\left(\sqrt{n+1}\rho_{n+1,m}-\sqrt{n}\rho_{n-1,m}\right) \nonumber\\
&+&E_{d}\left(\sqrt{m+1}\rho_{n,m+1}-\sqrt{m}\rho_{n,m-1}\right) \nonumber\\
&+&\kappa_{1}\left[\sqrt{(n+1)(m+1)}\rho_{n+1,m+1}-\frac{n+m}{2}\rho_{n,m}\right] \nonumber\\
&+&\kappa_{2}\left[\sqrt{(n+1)(n+2)}\sqrt{(m+1)(m+2)}\rho_{n+2,m+2}\right. \nonumber\\
&-&\left.\frac{n(n-1)+m(m-1)}{2}\rho_{n,m}\right],
\end{eqnarray}
where $\rho_{n,m}=\langle n|\hat{\rho}|m\rangle=\langle
m|\hat{\rho}|n\rangle^{*}$ is the matrix element of the density
operator $\hat{\rho}$. The diagonal element $\rho_{n,n}$ denotes the
occupying probability in the state $|n\rangle$, whereas the
off-diagonal element $\rho_{n,m}$ ($n\neq m$) represents the
coherence between the states $|n\rangle$ and $|m\rangle$.

Next, we approximate the oscillator modes by retaining the lowest
three oscillator levels, $n=0, 1, 2$, in Eq.\,(\ref{4}). In this
scenario, we have the equations of motion for the density operator
matrix element $\rho_{n,m}$ as follows:
\begin{eqnarray}
\frac{d\rho_{1,1}}{dt}&=&-\kappa_{1}\rho_{1,1}+2\kappa_{1}\rho_{2,2}+\sqrt{2}E_{d}\rho_{2,1}+\sqrt{2}E_{d}\rho_{1,2} \nonumber\\
&&-E_{d}\rho_{0,1}-E_{d}\rho_{1,0}, \label{5} \\
\frac{d\rho_{2,2}}{dt}&=&-2(\kappa_{1}+\kappa_{2})\rho_{2,2}-\sqrt{2}E_{d}\rho_{2,1}-\sqrt{2}E_{d}\rho_{1,2}, \label{6} \\
\frac{d\rho_{1,0}}{dt}&=&-\left(i\Delta+\frac{1}{2}\kappa_{1}\right)\rho_{1,0}+2E_{d}\rho_{1,1}+E_{d}\rho_{2,2} \nonumber\\
&&+\sqrt{2}\kappa_{1}\rho_{2,1}+\sqrt{2}E_{d}\rho_{2,0}-E_{d}, \label{7} \\
\frac{d\rho_{2,1}}{dt}&=&-\left(i\Delta+\frac{3}{2}\kappa_{1}+\kappa_{2}\right)\rho_{2,1}+\sqrt{2}E_{d}\left(\rho_{2,2}-\rho_{1,1}\right) \nonumber\\
&&-E_{d}\rho_{2,0}, \label{8} \\
\frac{d\rho_{2,0}}{dt}&=&-\left(2i\Delta+\kappa_{1}+\kappa_{2}\right)\rho_{2,0}+E_{d}\rho_{2,1} \nonumber\\
&&-\sqrt{2}E_{d}\rho_{1,0}, \label{9}
\end{eqnarray}
where $\rho_{0,0}=1-\rho_{1,1}-\rho_{2,2}$.

In the steady state limit, i.e, setting $d\rho_{n,m}/dt=0$ in the left-hand side of Eqs.\,(\ref{5})-(\ref{9}), we have the results
\begin{eqnarray}
&&-\kappa_{1}\rho_{1,1}+2\kappa_{1}\rho_{2,2}+\sqrt{2}E_{d}\rho_{2,1}+\sqrt{2}E_{d}\rho_{1,2}-E_{d}\rho_{0,1} \nonumber\\
&&-E_{d}\rho_{1,0}=0, \label{10} \\
&&-2(\kappa_{1}+\kappa_{2})\rho_{2,2}-\sqrt{2}E_{d}\rho_{2,1}-\sqrt{2}E_{d}\rho_{1,2}=0, \label{11} \\
&&-\left(i\Delta+\frac{1}{2}\kappa_{1}\right)\rho_{1,0}+2E_{d}\rho_{1,1}+E_{d}\rho_{2,2}+\sqrt{2}\kappa_{1}\rho_{2,1} \nonumber\\
&&+\sqrt{2}E_{d}\rho_{2,0}=E_{d}, \label{12} \\
&&-\left(i\Delta+\frac{3}{2}\kappa_{1}+\kappa_{2}\right)\rho_{2,1}+\sqrt{2}E_{d}\left(\rho_{2,2}-\rho_{1,1}\right) \nonumber\\
&&-E_{d}\rho_{2,0}=0, \label{13} \\
&&-\left(2i\Delta+\kappa_{1}+\kappa_{2}\right)\rho_{2,0}+E_{d}\rho_{2,1}-\sqrt{2}E_{d}\rho_{1,0}=0. \label{14}
\end{eqnarray}

This set of equations (\ref{10})-(\ref{14}) together with the
corresponding complex conjugations can be solved by first expressing
in the matrix form,
\begin{eqnarray}\label{15}
\mathbf{M}\mathbf{R}=\mathbf{V},
\end{eqnarray}
where the elements of the matrices are respectively presented by
$\mathbf{R}=(\rho_{1,1}, \rho_{2,2}, \rho_{1,0}, \rho_{0,1},
\rho_{2,1}, \rho_{1,2}, \rho_{2,0}, \rho_{0,2})^{T}$,
$\mathbf{V}=(0, 0, E_{d}, E_{d}, 0, 0, 0, 0)^{T}$, and
\begin{widetext}
\begin{eqnarray}
\mathbf{M}=
\left(
\begin{array}{cccccccc}
-\kappa_{1} & 2\kappa_{2} & -E_{d} & -E_{d} & \sqrt{2}E_{d} & \sqrt{2}E_{d} & 0 & 0 \\
0 & -2(\kappa_{1}+\kappa_{2}) & 0 & 0 & -\sqrt{2}E_{d} & -\sqrt{2}E_{d} & 0 & 0 \\
2E_{d} & E_{d} & -i\Delta-\frac{1}{2}\kappa_{1} & 0 & \sqrt{2}\kappa_{1} & 0 & \sqrt{2}E_{d} & 0 \\
2E_{d} & E_{d} & 0 & i\Delta-\frac{1}{2}\kappa_{1} & 0 & \sqrt{2}\kappa_{1} & 0 & \sqrt{2}E_{d} \\
-\sqrt{2}E_{d} & \sqrt{2}E_{d} & 0 & 0 & -i\Delta-\frac{3}{2}\kappa_{1}-\kappa_{2} & 0 & -E_{d} & 0 \\
-\sqrt{2}E_{d} & \sqrt{2}E_{d} & 0 & 0 & 0 & i\Delta-\frac{3}{2}\kappa_{1}-\kappa_{2} & 0 & -E_{d} \\
0 & 0 & -\sqrt{2}E_{d} & 0 & E_{d} & 0 & -2i\Delta-\kappa_{1}-\kappa_{2} & 0 \\
0 & 0 & 0 & -\sqrt{2}E_{d} & 0 & E_{d} & 0 & 2i\Delta-\kappa_{1}-\kappa_{2}  \\
\end{array}
\right). \nonumber
\end{eqnarray}
\end{widetext}

The solutions of $\mathbf{R}$ are given by
$\mathbf{R}=\mathbf{M}^{-1}\mathbf{V}$ with $\mathbf{M}^{-1}$
denoting the inverse matrix of $\mathbf{M}$, and here
$\mathbf{M}^{-1}$ is too lengthy and uninspiring to be displayed. In
order to capture the main physics, we consider some particular cases
of interest and derive closed-form expressions of $\mathbf{R}$.
First, the driving is set on resonance with the vdP oscillator,
i.e., $\Delta=0$ for simplicity. Second, the condition of weak
external driving holds, i.e., $E_{d}\ll\kappa_{1}, \kappa_{2}$, in
order to guarantee the validity of the three-level model. With these
assumptions discussed above, $\rho_{1,1}$ and $\rho_{2,2}$ we need
in the solutions of $\mathbf{R}$ [see Eq.\,(\ref{20a}) later for
clarity], to the quartic order in the variables $E_{d}/\kappa_{1}$
and $E_{d}/\kappa_{2}$ after tedious calculations, are
\begin{widetext}
\begin{eqnarray}
\rho_{1,1}&=&\left[2+\frac{\kappa_{1}^{2}}{4E_{d}^{2}}
+\frac{2\kappa_{1}\kappa_{2}-2\kappa_{1}^{2}}{(\kappa_{1}+\kappa_{2})(3\kappa_{1}+2\kappa_{2})}
-\frac{4\kappa_{1}}{3\kappa_{1}+2\kappa_{2}}+\frac{\kappa_{1}}{\kappa_{1}+\kappa_{2}}
+\frac{8\kappa_{2}E_{d}^{2}}{(\kappa_{1}+\kappa_{2})^{2}(3\kappa_{1}+2\kappa_{2})}\right]^{-1}, \label{16} \\
\rho_{2,2}&=&\left[\frac{4E_{d}^{2}}{(\kappa_{1}+\kappa_{2})(3\kappa_{1}+2\kappa_{2})}
+\frac{2\kappa_{1}E_{d}^{2}}{(\kappa_{1}+\kappa_{2})^{2}(3\kappa_{1}+2\kappa_{2})}\right]\rho_{1,1}. \label{17}
\end{eqnarray}
\end{widetext}

\subsection{Analytical expressions for the second-order correlation function} \label{IVB}
According to $\hat{a}^{\dag}\hat{a}|n\rangle=n|n\rangle$ and
$\hat{a}^{\dag2}\hat{a}^{2}|n\rangle=n(n-1)|n\rangle$, we have
$\langle n|\hat{\rho}\hat{a}^{\dag}\hat{a}|n\rangle=n\rho_{n,n}$ and
$\langle
n|\hat{\rho}\hat{a}^{\dag2}\hat{a}^{2}|n\rangle=n(n-1)\rho_{n,n}$.
So, the zero-time-delay second-order correlation function
$g^{(2)}(0)$ can be recast in the form
\begin{eqnarray}\label{18}
g^{(2)}(0)&=&\frac{\sum\limits_{n}n(n-1)\rho_{n,n}}{\left(\sum\limits_{n}n\rho_{n,n}\right)^{2}} \nonumber \\
&=&\frac{2\rho_{2,2}+6\rho_{3,3}+\cdots}{(\rho_{1,1}+2\rho_{2,2}+\cdots)^{2}},
\end{eqnarray}
where the sum index $n=0, 1, 2, \cdots$. Obviously, nonclassical
antibunching (sub-Poissonian statistics) requires that the numerator
of Eq.\,(\ref{18}) is smaller than its denominator. For example,
this can be achieved either by maximizing $\rho_{1,1}$ or decreasing
$\rho_{2,2}$. Note that, the component of $n=0$ need not be
considered owing to the vanishing value.

Under the weak driving condition where $\rho_{n,n}$ vanishes quickly with $n$, we have the relationship
\begin{eqnarray}\label{19}
\rho_{0,0}\gg\rho_{1,1}\gg\rho_{2,2}\gg\rho_{3,3}\gg\cdots.
\end{eqnarray}
In view of this, by the truncation, the $g^{(2)}(0)$ function from
Eq.\,(\ref{18}) can be approximately reduced into the simple
expression
\begin{eqnarray}\label{20a}
g^{(2)}(0)\simeq\frac{2\rho_{2,2}}{\rho_{1,1}^{2}},
\end{eqnarray}
or
\begin{eqnarray}\label{20b}
g^{(2)}(0)&\simeq&\left[\frac{8E_{d}^{2}}{(\kappa_{1}+\kappa_{2})(3\kappa_{1}+2\kappa_{2})}\right. \nonumber\\
&&\left.+\frac{4\kappa_{1}E_{d}^{2}}{(\kappa_{1}+\kappa_{2})^{2}(3\kappa_{1}+2\kappa_{2})}\right]\times\frac{1}{\rho_{1,1}},
\end{eqnarray}
where the two desired matrix elements $\rho_{2,2}$ and $\rho_{1,1}$ are yielded by Eqs.\,(\ref{16}) and (\ref{17}).

As an aside, we note that when the two-phonon loss is much stronger
than the single-phonon loss ($\kappa_{2}\gg\kappa_{1}$), retaining
the lowest order in $E_{d}/\kappa_{1}$, $E_{d}/\kappa_{2}$ and
$\kappa_{1}/\kappa_{2}$, the $g^{(2)}(0)$ function from
Eq.\,(\ref{20b}) can be further reduced into a compact form
\begin{eqnarray}\label{21}
g^{(2)}(0)\simeq\left(\frac{\kappa_{1}}{\kappa_{2}}\right)^{2}\sim0,
\end{eqnarray}
which coincides with the numerical simulation based on the full
quantum master equation (\ref{2}) in the aforementioned parameter
range, as can be found in Fig.\,\ref{fig1} below.

\subsection{Three-oscillator-level approximation via a Schr\"{o}dinger equation approach} \label{IVC}
From another point of view, now we consider the Schr\"{o}dinger
equation approach to analytically calculate the second-order
correlation function. Likewise, under the condition that the
external driving laser field is very weak (i.e.,
$E_{d}\ll\kappa_{1}, \kappa_{2}$), the aforementioned excitation
number $n$ of the vdP oscillator is no more than two (i.e., $n\leq2$
corresponding to the so-called three-oscillator-level truncation).
In this scenario, the quantum state of the vdP oscillator system can
be written as
\begin{eqnarray}\label{22}
|\psi\rangle=c_{0}|0\rangle+c_{1}|1\rangle+c_{2}|2\rangle,
\end{eqnarray}
where the coefficient $c_{n}$ represents the amplitude of the
corresponding quantum state $|n\rangle$ ($n=0, 1, 2$).

In order to achieve the coefficient $c_{n}$ above, we need to solve
the following Schr\"{o}dinger equation
\begin{eqnarray}\label{23}
i\frac{\partial|\psi\rangle}{\partial
t}=\hat{\widetilde{H}}_{osc}|\psi\rangle,
\end{eqnarray}
where $\hat{\widetilde{H}}_{osc}$ is the modified non-Hermitian
Hamiltonian of the system including phenomenologically both the
linear single-phonon loss ($\kappa_{1}$) and nonlinear two-phonon
loss ($\kappa_{2}$) terms, with the form
\begin{eqnarray}\label{24}
\hat{\widetilde{H}}_{osc}=-i\frac{1}{2}\kappa_{1}\hat{a}^{\dag}\hat{a}
-i\frac{1}{2}\kappa_{2}\hat{a}^{\dag2}\hat{a}^{2}+\Delta\hat{a}^{\dag}\hat{a}+iE_{d}(\hat{a}-\hat{a}^{\dag}).
\nonumber \\
\end{eqnarray}
Above, the first and second terms added phenomenologically can be
well understood from the Lindblad operators in Eq.\,(\ref{2}). Since
we consider the weak driving limit, the vdP oscillator is rarely in
the excited states and therefore the contributions of the
$2\hat{\mathcal{O}}\hat{\rho}\hat{\mathcal{O}}^{\dag}$ terms
appearing in the master equation can be safely omitted. This just is
equivalent to acquiring the effective Hamiltonian (\ref{24}).

According to Eq.\,(\ref{23}), we can directly obtain a set of
coupled algebraic equations about the coefficient $c_{n}$ as follows
\begin{eqnarray}
\frac{\partial c_{0}}{\partial t}&=&E_{d}c_{1}, \label{25a} \\
\frac{\partial c_{1}}{\partial t}&=&-i\Delta
c_{1}-\frac{1}{2}\kappa_{1}c_{1}-E_{d}c_{0}+\sqrt{2}E_{d}c_{2}, \label{25b} \\
\frac{\partial c_{2}}{\partial t}&=&-2i\Delta
c_{2}-\kappa_{1}c_{2}-\kappa_{2}c_{2}-\sqrt{2}E_{d}c_{1}.
\label{25c}
\end{eqnarray}

In the limit of the above-mentioned weak-driving, we can take
$c_{0}\simeq1$ [thus Eq.\,(\ref{25a}) can be dropped] and neglect
the third-order term $\sqrt{2}E_{d}c_{2}$ (we only retain the terms
up to the second order) in Eq.\,(\ref{25b}). Under the steady-state
situation, $\partial|c_{n}\rangle/\partial t=0$, we have the results
\begin{eqnarray}
&&-i\Delta c_{1}-\frac{1}{2}\kappa_{1}c_{1}-E_{d}=0, \label{26a} \\
&&-2i\Delta
c_{2}-\kappa_{1}c_{2}-\kappa_{2}c_{2}-\sqrt{2}E_{d}c_{1}=0.
\label{26b}
\end{eqnarray}

After some straightforward calculations, the solutions to
Eqs.\,(\ref{26a}) and (\ref{26b}) for the coefficients $c_{1}$ and
$c_{2}$ are found to be
\begin{eqnarray}
c_{1}&=&-\frac{2E_{d}}{2i\Delta+\kappa_{1}}, \label{27a} \\
c_{2}&=&-\frac{\sqrt{2}E_{d}c_{1}}{2i\Delta+\kappa_{1}+\kappa_{2}},
\label{27b}
\end{eqnarray}
which are closely related to the matrix elements $\rho_{1,1}$ and
$\rho_{2,2}$ via the relationships $\rho_{1,1}=|c_{1}|^{2}$ and
$\rho_{2,2}=|c_{2}|^{2}$. So, based on Eq.\,(\ref{20a}), we can
derive a closed-form expression for the $g^{(2)}(0)$ function as
\begin{eqnarray}\label{28}
g^{(2)}(0)\simeq\frac{\kappa_{1}^{2}+4\Delta^{2}}{(\kappa_{1}+\kappa_{2})^{2}+4\Delta^{2}},
\end{eqnarray}
where for the significantly large detuning $\Delta$, $g^{(2)}(0)$
approaches the value of $1$. For the resonance driving $\Delta=0$,
the $g^{(2)}(0)$ function is further simplified to
$g^{(2)}(0)\simeq\kappa_{1}^{2}/(\kappa_{1}+\kappa_{2})^{2}$. In
this scenario of $\Delta=0$, for large $\kappa_{2}$, $g^{(2)}(0)$
tends to the value of $0$, while for small $\kappa_{2}$,
$g^{(2)}(0)$ tends to the value of $1$. Theses results are in good
agreement with the numerical simulations based on the full quantum
master equation (\ref{2}) in Figs.\,\ref{fig1}(a) and \ref{fig1}(b)
later.

Physically, the zero value of $g^{(2)}(0)$ can be understood from
the $\kappa_{2}/\kappa_{1}$ ratio determining the structure of the
discrete energy level in the quantum circumstance: the quantum vdP
oscillator is favorable to be restricted to the lowest Fock states
as the nonlinear two-phonon loss $\kappa_{2}$ increases, which has
been pointed out in Refs.\,\cite{I-79, I-80}. In the limit of
$\kappa_{2}\gg\kappa_{1}$, only the two lowest Fock states
$|0\rangle$ and $|1\rangle$ are occupied. In this quantum regime,
the nonlinearity of the dissipation prevents reaching the two phonon
state, that is to say, the transition pathway from Fock states
$|1\rangle$ to $|2\rangle$ is virtually forbidden, so realizing
phonon blockade is possible.

Finally, it is worth pointing out that, via comparing the two-type
treatments [(i) the master equation approach and (ii) the
Schr\"{o}dinger equation approach], it is clearly shown that the
Schr\"{o}dinger equation approach is more robust and
straightforward. Nevertheless, its deficiency is that the
dissipations of the system are phenomenologically added.

\section{Result analyses and discussions} \label{V}
\begin{figure*}[htb]
\centerline{\includegraphics[width=8.5cm]{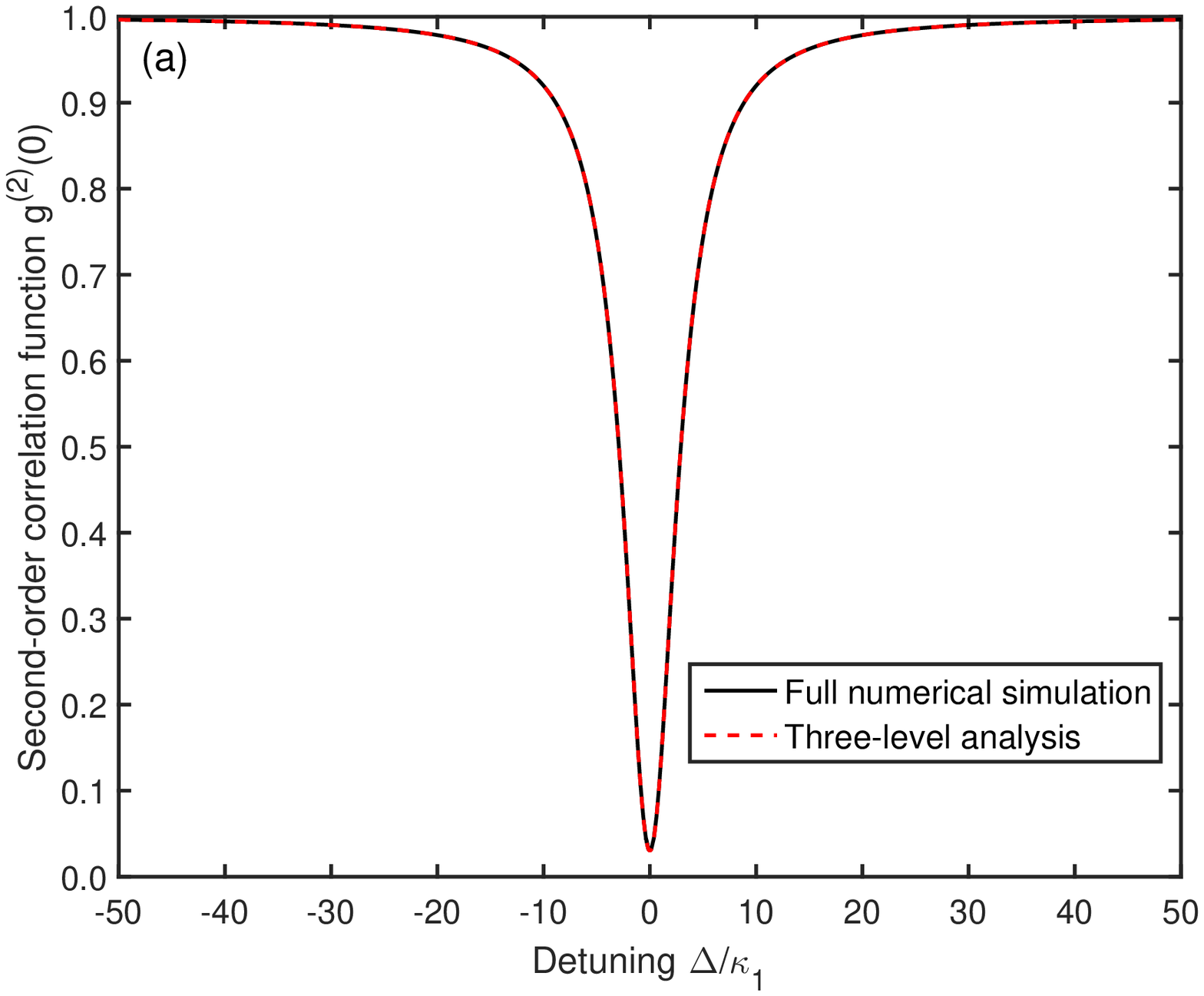}
\includegraphics[width=8.5cm]{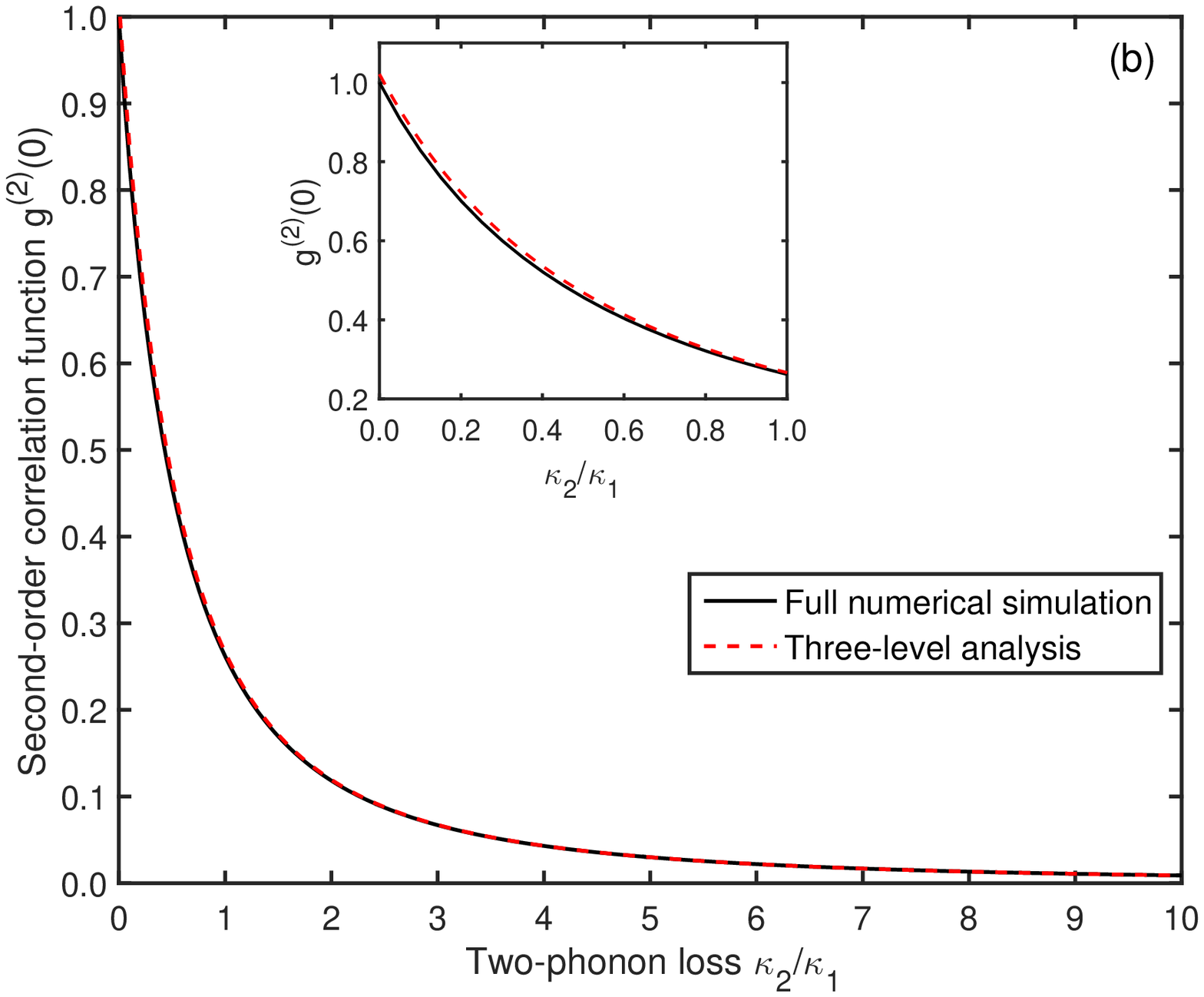}}
\centerline{\includegraphics[width=8.5cm]{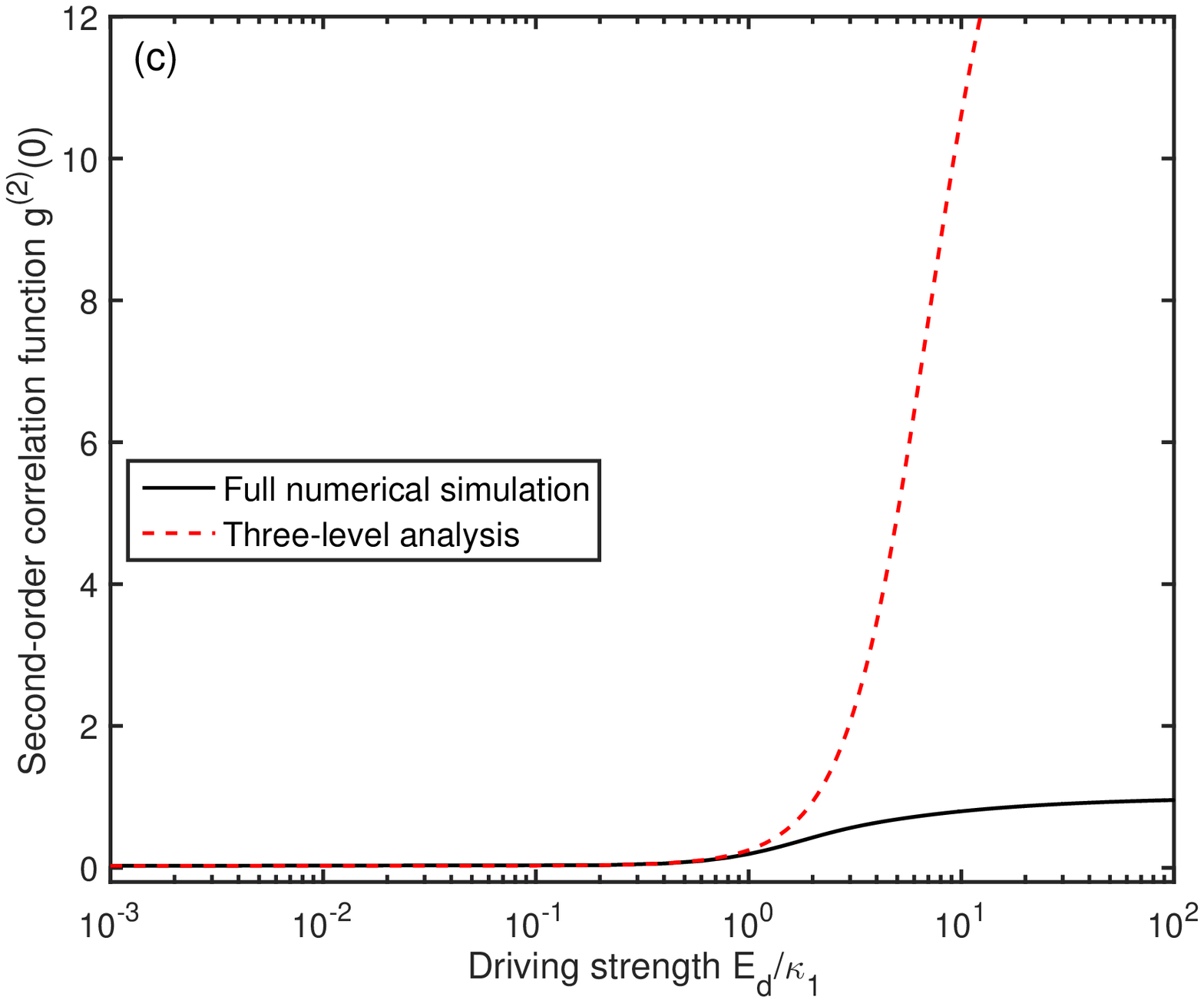}}
\caption{\label{fig1} (a) Normalized zero-delay second-order
correlation function $g^{(2)}(0)$ of the vdP oscillator plotted as a
function of the dimensionless detuning $\Delta/\kappa_{1}$ for the
two-phonon loss rate $\kappa_{2}/\kappa_{1}=5$ and the driving
strength $E_{d}/\kappa_{1}=0.1$. (b) Dependence of zero-delay
second-order correlation function $g^{(2)}(0)$ on the dimensionless
two-phonon loss rate $\kappa_{2}/\kappa_{1}$, plotted for the zero
detuning $\Delta/\kappa_{1}=0$ and the driving strength
$E_{d}/\kappa_{1}=0.1$. The inset shows a zoom-in view of the
$\kappa_{2}<\kappa_{1}$ region. (c) Dependence of zero-delay
second-order correlation function $g^{(2)}(0)$ on the dimensionless
driving strength $E_{d}/\kappa_{1}$ for changing logarithmically
from $0.001$ to $100$, plotted for the zero detuning
$\Delta/\kappa_{1}=0$ and the two-phonon loss rate
$\kappa_{2}/\kappa_{1}=5$. In panels (a)-(c), the black solid lines
denote full numerical simulations of a quantum master equation from
Eq.\,(\ref{2}), whereas the red dashed lines are analytical
calculations of a three-oscillator-level model from
Eq.\,(\ref{15}).}
\end{figure*}

Figure \ref{fig1} shows the steady-state value of the normalized
second-order correlation function at zero-time delay, $g^{(2)}(0)$,
as a function of the detuning $\delta$ [panel (a)], the two-phonon
loss $\kappa_{2}$ [panel (b)], and the driving strength $E_{d}$
[panel (c)], respectively. All plotted parameters are in units of
the single-phonon loss $\kappa_{1}$ and thus are dimensionless. The
black solid lines represent full numerical simulations from the
quantum master equation (\ref{1}), while the red dashed lines are
analytical calculations predicted by Eq.\,(\ref{15}) from the
three-oscillator-level approximation. To be specific, it can be seen
from Fig.\,\ref{fig1}(a) that the profile of the second-order
correlation function $g^{(2)}(0)$ is symmetric with respect to the
detuning $\Delta$. There is a marked antibunching dip at $\Delta=0$,
corresponding to a minimum value of $g^{(2)}(0)\simeq0.03$ in good
agreement with Eq.\,(\ref{28}). Increasing $|\Delta|$ leads to an
increase in $g^{(2)}(0)$, degrading the degree of antibunching. For
significantly large $|\Delta|$, $g^{(2)}(0)$ approaches a saturation
value of unity, indicating coherent light emission. Numerical
simulations matching analytical calculations show that the strong
phonon antibunching can be achieved by utilizing the typical
parameter values, e.g., $\kappa_{2}=5\kappa_{1}$ and
$E_{d}=0.1\kappa_{1}$ given in Fig.\,\ref{fig1}(a), not specific
values from any one experiment. In order to understand the physics
behind this antibunching effect, intuitively, we look at the role of
the $\kappa_{2}$ parameter, as will be discussed later.

\begin{figure}[htb]
\centerline{\includegraphics[width=9.7cm]{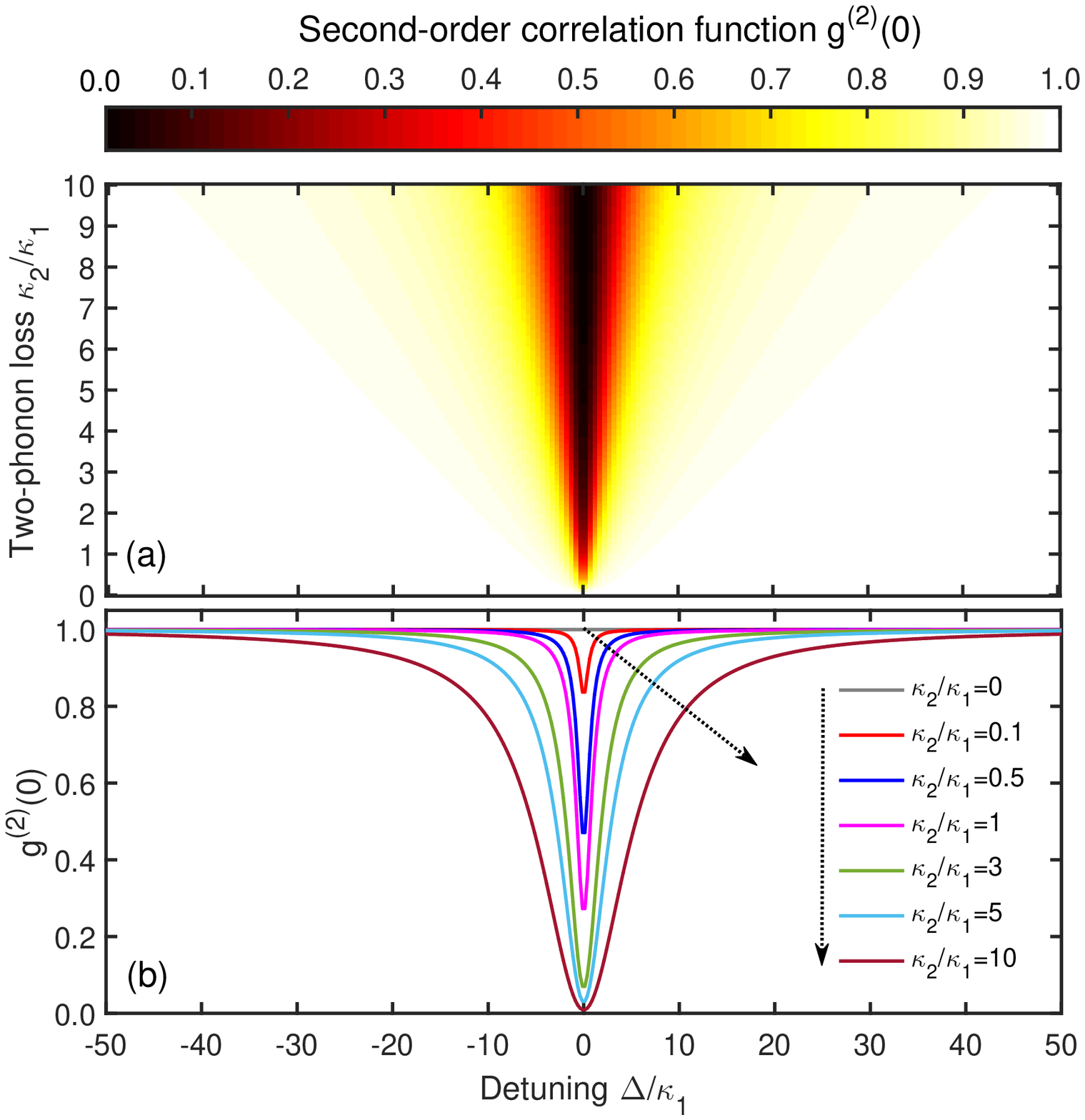}}
\caption{\label{fig2} (a) 2D contour plot of normalized zero-delay
second-order correlation function $g^{(2)}(0)$ versus the detuning
$\Delta/\kappa_{1}$ and the two-phonon loss rate
$\kappa_{2}/\kappa_{1}$ for the driving strength
$E_{d}/\kappa_{1}=0.1$. (b) Cross section of (a) along the $\Delta$
axis, for seven fixed sets of the two-phonon loss rate
$\kappa_{2}/\kappa_{1}=0$, $0.1$, $0.5$, $1$, $3$, $5$, and $10$.}
\end{figure}

\begin{figure}[htb]
\centerline{\includegraphics[width=9.7cm]{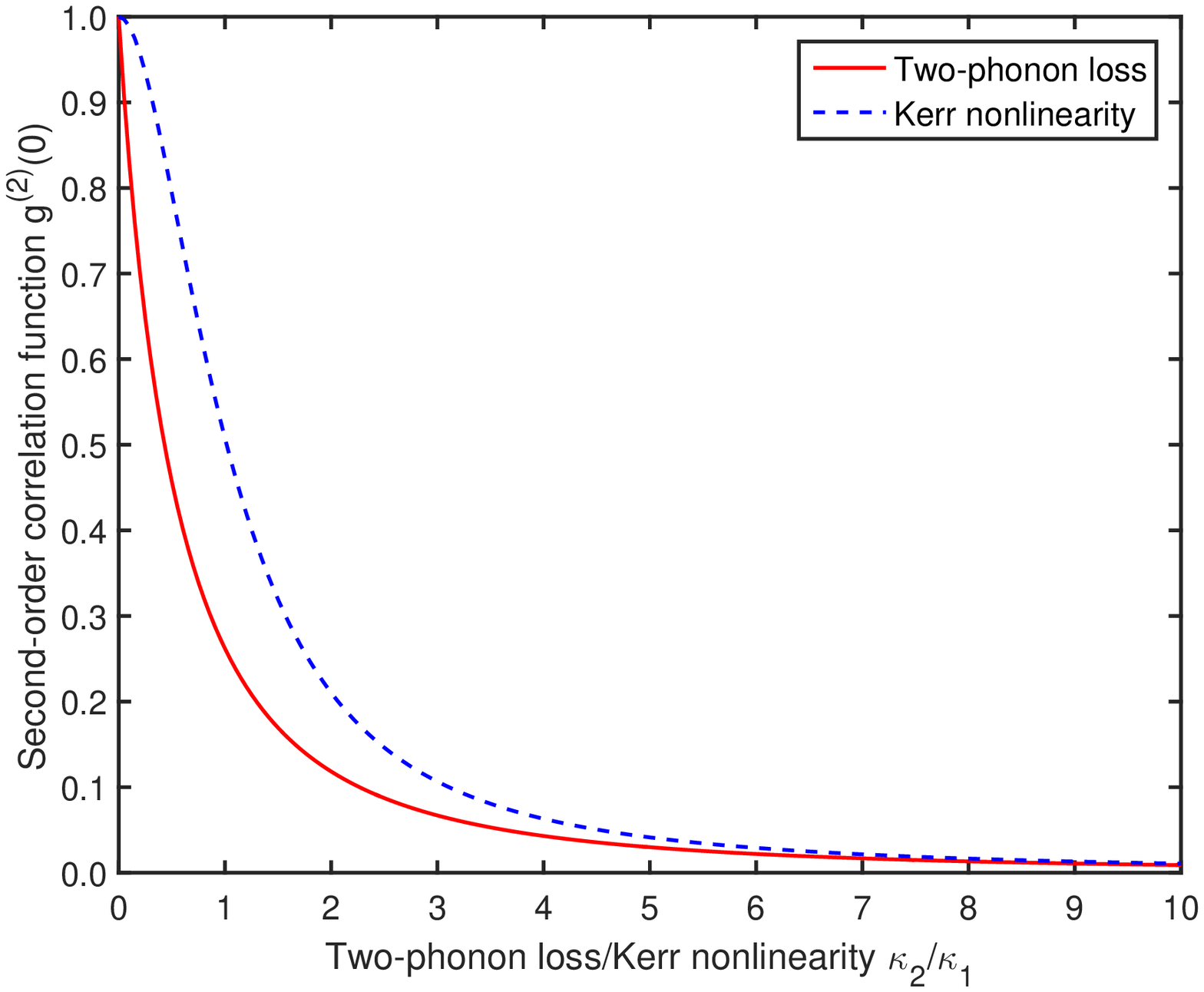}}
\caption{\label{fig3} Comparison of the normalized second-order
correlation function $g^{(2)}(0)$ obtained via the two-phonon loss
in the vdP oscillator to that obtained via the Kerr nonlinearity in
the resonator which only includes a single-phonon loss. Here the
resulting effective Hamiltonian and the Lindblad master equation of
the Kerr nonlinear resonator, respectively, are yielded by
$\hat{H}_{KN}=\Delta\hat{a}^{\dag}\hat{a}+\frac{1}{2}\kappa_{2}\hat{a}^{\dag2}\hat{a}^{2}+iE_{d}(\hat{a}-\hat{a}^{\dag})$
and $d\hat{\rho}/dt=-i[\hat{H}_{KN},
\hat{\rho}]+\kappa_{1}D(\hat{a})\hat{\rho}$ in a rotating frame with
respect to $\hat{H}_{0}=\omega_{d}\hat{a}^{\dag}\hat{a}$. See main
text about the vdP oscillator for more details. The other system
parameters are set as $E_{d}/\kappa_{1}=0.1$ and
$\Delta/\kappa_{1}=0$.}
\end{figure}

As illustrated in Fig.\,\ref{fig1}(b), when the two-phonon loss
$\kappa_{2}$ is set to zero, $g^{(2)}(0)$ is equal to $1$,
suggesting coherent light emitted from the vdP oscillator. Again, we
can clearly see that $g^{(2)}(0)$ monotonically and rapidly
decreases from $1$ as $\kappa_{2}$ increases, asymptotically
approaching zero for large $\kappa_{2}$. For
$\kappa_{2}\sim0.4\kappa_{1}$, $g^{(2)}(0)$ can reach $0.5$. The
value $g^{(2)}(0)<0.5$ is considered as an upper bound for
single-phonon emission \cite{I-35b,I-62,I-66,V-4}. As a consequence,
the criterion for single-phonon emission can hold when
$\kappa_{2}>0.4\kappa_{1}$. Such dissipative processes can be
engineered in current experimental devices \cite{VI-1,VI-2,VI-15},
as will be illustrated in Sec.\,\ref{VI} below. Note that, in
Fig.\,\ref{fig1}(b), we keep the parameters $\Delta=0$ and
$E_{d}=0.1\kappa_{1}$ fixed. In this scenario, the analytical
$g^{(2)}(0)$ [black solid line in Fig.\,\ref{fig1}(b)] agrees very
well with the numerically exact results [red dashed line in
Fig.\,\ref{fig1}(b)]. Overall, the two-phonon loss dependence of
phonon emission statistics shows that the degree of phonon
antibunching can be kept high even for $\kappa_{2}<\kappa_{1}$.

In light of the above analyses from Figs.\,\ref{fig1}(a) and (b), we
can conclude that the validity of such a three-oscillator-level
approximation is not dependent on both the $\Delta$ and $\kappa_{2}$
parameters. What is the breakdown of this three-oscillator-level
approximation? To help answer this question, in Fig.\,\ref{fig1}(c)
we displays a typical case for varying the driving strength $E_{d}$
while keeping $\Delta=0$ and $\kappa_{2}=5\kappa_{1}$ fixed. Looking
closer, we see that for small driving strengths in the range
$E_{d}\leq\kappa_{1}$, this three-level analysis closely matches the
result of the master equation simulation shown in
Fig.\,\ref{fig1}(c). However, when the ratio $E_{d}/\kappa_{1}$ is
larger than $1$, the three-level approximation completely fails to
account for $g^{(2)}(0)$. With this fact, we notice from the black
solid line of Fig.\,\ref{fig1}(c) that, as $E_{d}$ increases
logarithmically from $0.001$ to $100$, $g^{(2)}(0)$ grows from a
constant value close to zero, corresponding to single-phonon
emission ($E_{d}\leq\kappa_{1}$), to an unity saturation value
indicating coherent light emission ($E_{d}>80\kappa_{1}$). It is
worth pointing out that strong antibunching [$g^{(2)}(0)\sim0$]
features a long plateau about $E_{d}$.

\begin{figure*}[htb]
\centerline{\includegraphics[width=8.5cm]{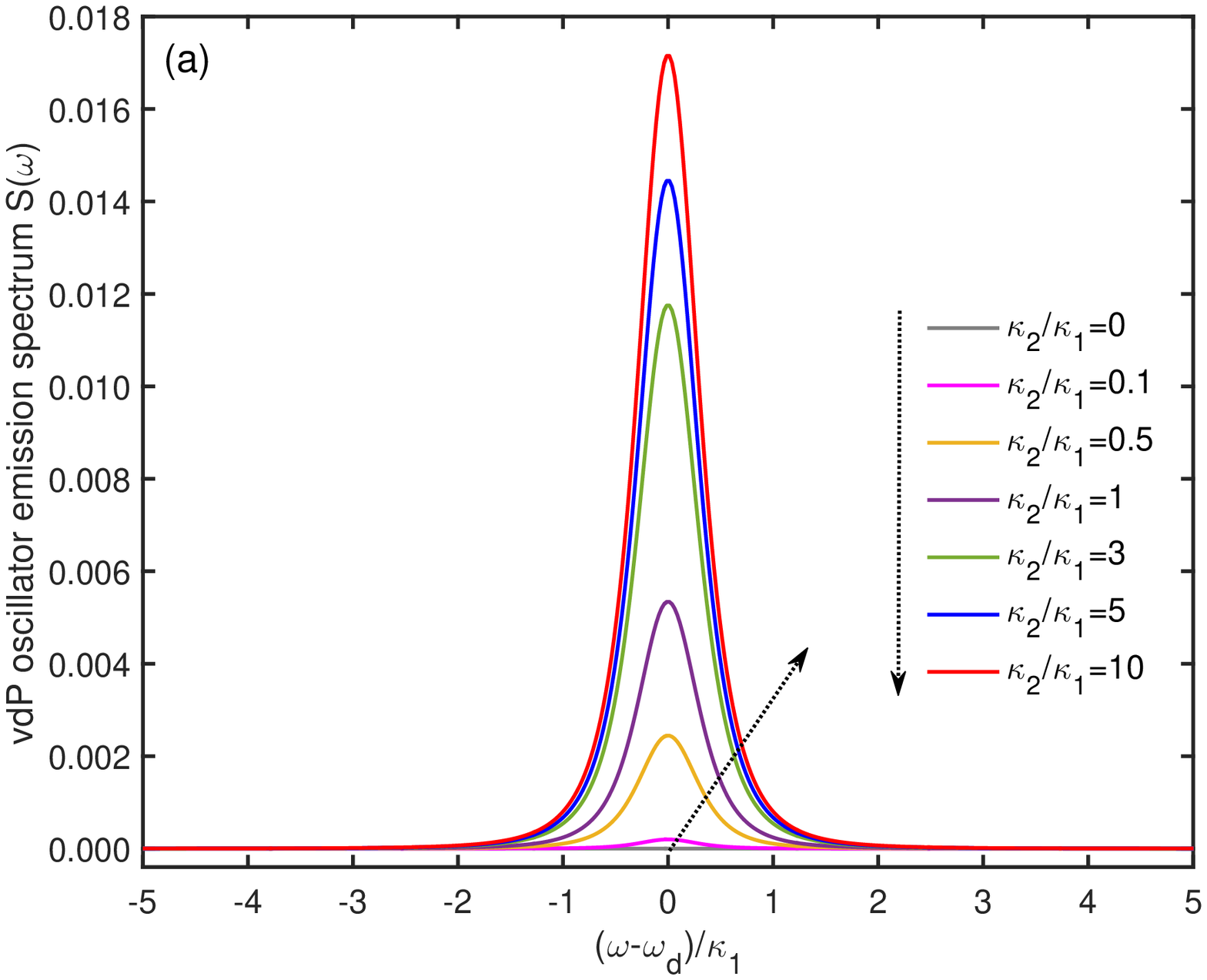}
\includegraphics[width=8.5cm]{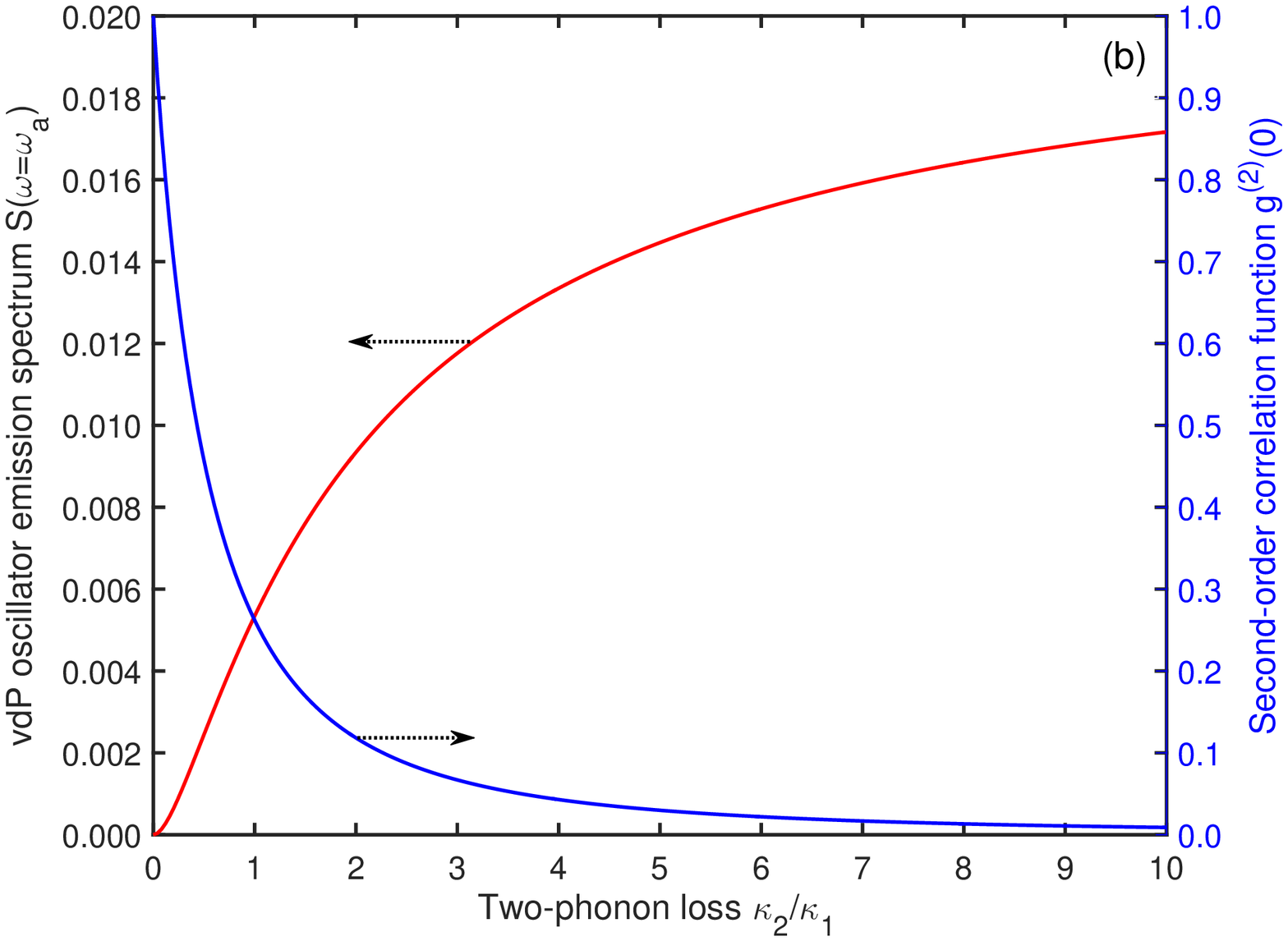}}
\centerline{\includegraphics[width=8.5cm]{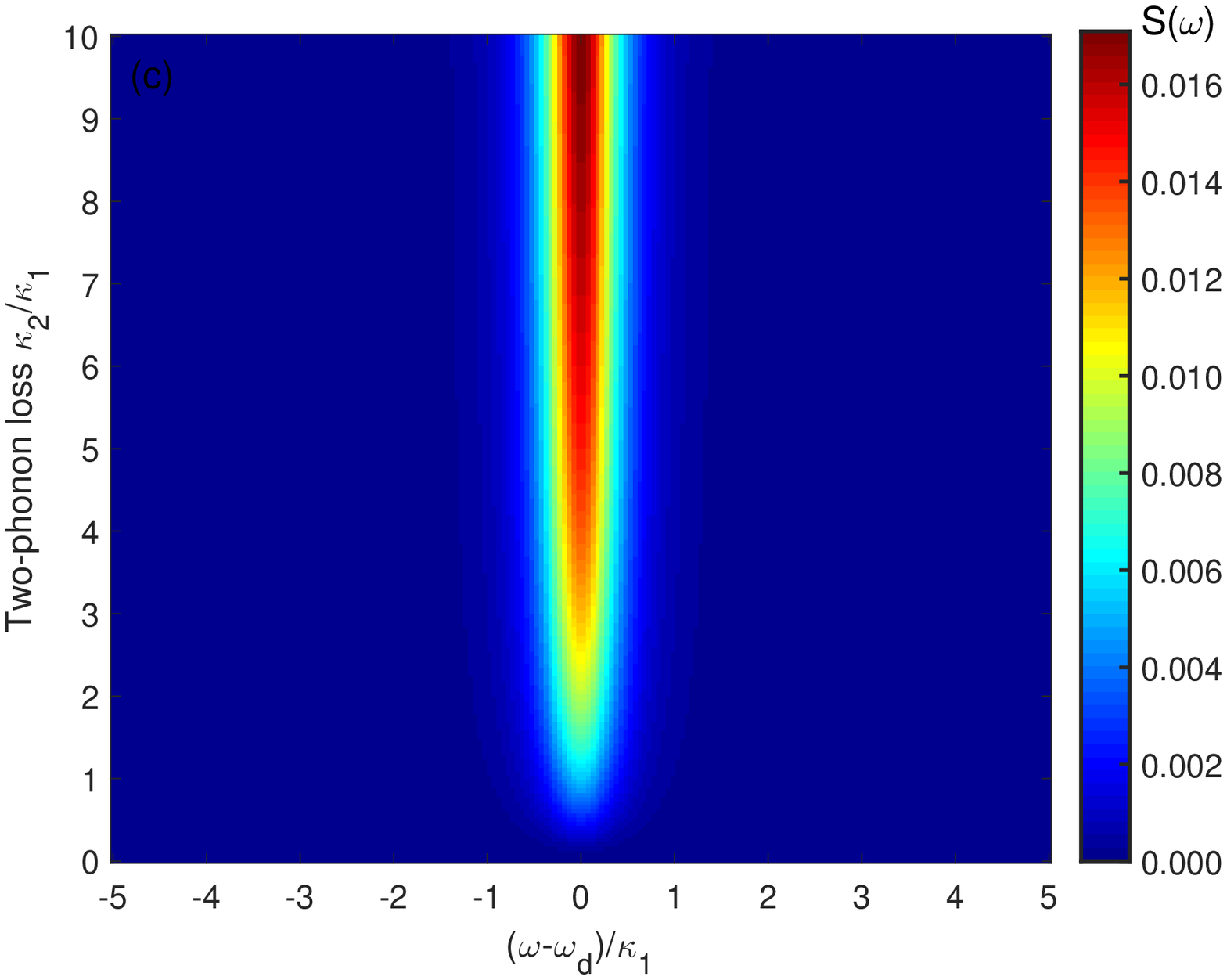}}
\caption{\label{fig4} (a) Phonon emission spectrum $S(\omega)$ in
arbitrary units, i.e., PSD as a function of the dimensionless
frequency $(\omega-\omega_{d})/\kappa_{1}$ for a resonantly driven
vdP oscillator in a frame rotating at the driving frequency
$\omega_{d}$. The other system parameters are chosen as
$E_{d}/\kappa_{1}=0.1$ and $\Delta/\kappa_{1}=0$, respectively. (b)
Value of the spectrum $S(\omega=\omega_{a})$ at the vdP oscillator
frequency $\omega_{a}$ as a function of the two-phonon loss rate
$\kappa_{2}/\kappa_{1}$ on the left-hand axis. Zero-delay
second-order correlation function $g^{(2)}(0)$ as a function of the
two-phonon loss rate $\kappa_{2}/\kappa_{1}$ on the right-hand axis.
The other system parameters are set as $E_{d}/\kappa_{1}=0.1$ and
$\Delta/\kappa_{1}=0$. (c) 2D contour plot of vdP oscillator
emission spectrum $S(\omega)$ (in arbitrary units) versus the
frequency $(\omega-\omega_{d})/\kappa_{1}$ and the two-phonon loss
rate $\kappa_{2}/\kappa_{1}$ for $E_{d}/\kappa_{1}=0.1$ and
$\Delta/\kappa_{1}=0$.}
\end{figure*}

\begin{figure}[htb]
\centerline{\includegraphics[width=9.5cm]{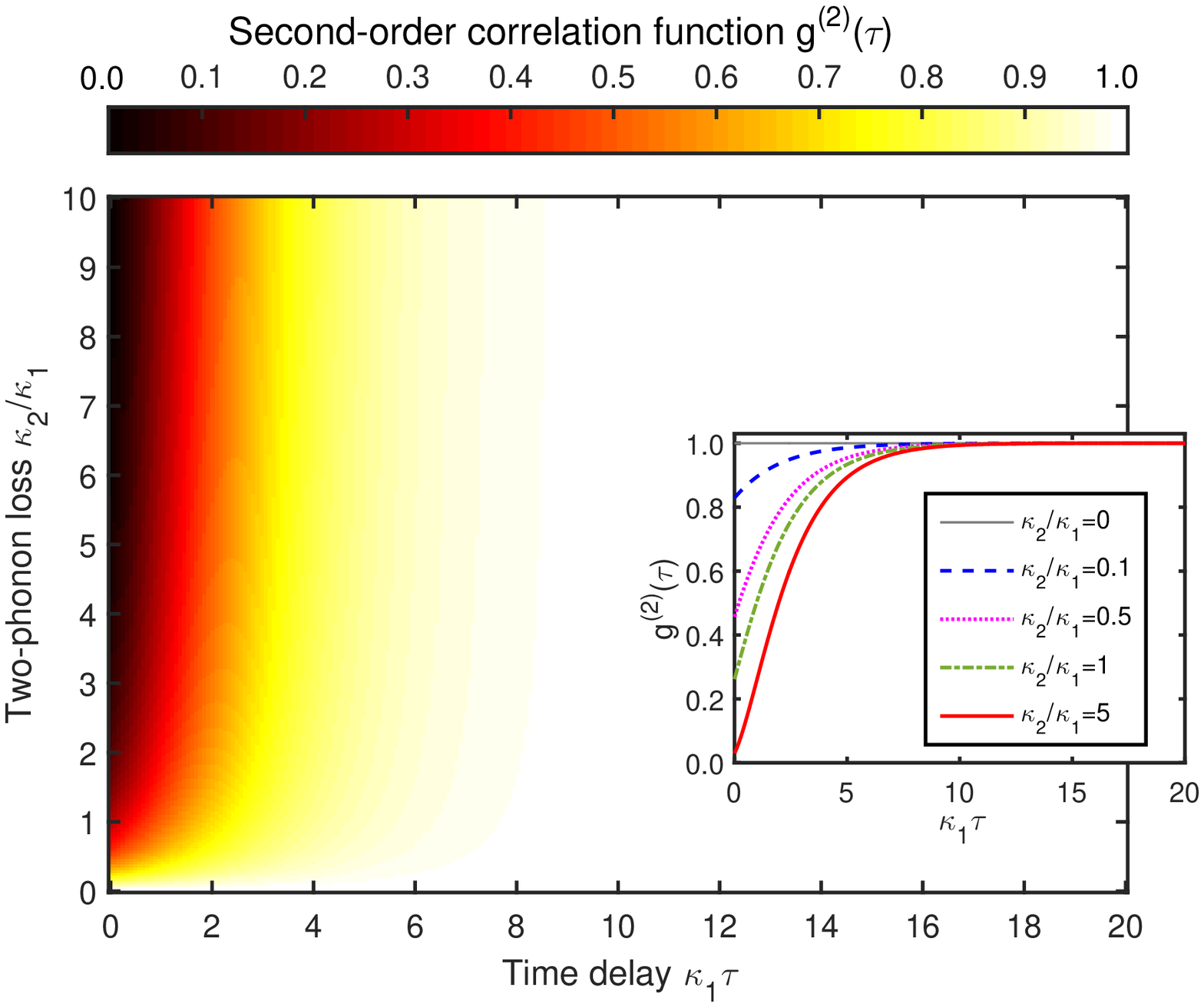}}
\caption{\label{fig5} 2D contour plot of normalized second-order
correlation function $g^{(2)}(\tau)$ with time delay $\tau$ as a
function of the time delay $\kappa_{1}\tau$ and the two-phonon loss
rate $\kappa_{2}/\kappa_{1}$ for the driving strength
$E_{d}/\kappa_{1}=0.1$. The inset show the cross section of the main
plot along the lateral axis, i.e., the time-delay dependence of the
second-order correlation function $g^{(2)}(\tau)$, with the five
different values of the two-phonon loss rate
$\kappa_{2}/\kappa_{1}=0$, $0.1$, $0.5$, $1$, and $5$.}
\end{figure}

To gain further insight, Figure \ref{fig2}(a) displays the
color-scale maps of the normalized second-order correlation function
$g^{(2)}(0)$ as a function of both the detuning $\Delta$ and the
two-phonon loss $\kappa_{2}$. Figure \ref{fig2}(b) shows the cuts
through the horizontal $\Delta$-axis for seven different values of
$\kappa_{2}$. It follows that the antibunching window width
increases with increasing $\kappa_{2}$. Moreover, the degree of the
antibunching is more pronounced as $\kappa_{2}$ is increased,
whether the applied driving is set on resonance with the vdP
oscillator or not. A natural question is, then, why the dissipation
$\kappa_{2}$ plays an important role in generating phonon
antibunching? As it is well known, the phonon dissipation offers a
major source of decoherence \cite{V-5,V-6,V-7,V-10}, giving rise to
the expectation that the dissipation should always weaken the
antibunching. Physically, this is mainly because the two terms
($\frac{1}{2}\kappa_{2}a^{\dag2}a^{\dag2} \rho$ and
$\frac{1}{2}\kappa_{2}\rho a^{\dag2}a^{\dag2}$) in the two-phonon
dissipation expression of Eq.\,(\ref{1}) act as the Kerr
nonlinearity $\chi^{(3)}$ (see, for instance,
Refs.\,\cite{V-8,V-9,VI-7}, and references therein). As an example,
the dynamical evolution equations of the key field operator in the
respective system are analogous; specifically, one is
$\partial\hat{a}/\partial
t=-i\Delta\hat{a}-\frac{1}{2}\kappa_{1}\hat{a}-\kappa_{2}\hat{a}^{\dag}\hat{a}\hat{a}-E_{d}$
for the vdP oscillator, the other is $\partial\hat{a}/\partial
t=-i\Delta\hat{a}-\frac{1}{2}\kappa_{1}\hat{a}-i\chi^{(3)}\hat{a}^{\dag}\hat{a}\hat{a}-E_{d}$
for the Kerr nonlinear resonator. However, the present vdP
oscillator for realizing the antibunching outperforms the Kerr
nonlinear resonator when setting $\chi^{(3)}=\kappa_{2}$ and the
other common parameters are the same as in Fig.\,\ref{fig1}(a), as
verified from Fig.\,\ref{fig3}. A striking feature of the results of
Fig.\,\ref{fig3} is the fact that, for an identical $\kappa_{2}$,
the degree of the phonon antibunching for the vdP oscillator is much
smaller than that for the Kerr nonlinear resonator. Thus, the
antibunching phenomenon is more pronounced relative to the Kerr
nonlinear resonator.

In the following, we elaborate on the amount of phonon emission
through the vdP oscillator corresponding to the driving used in
Figs.\,\ref{fig1}(a) and (b), as we collect the fluorescence
primarily from the oscillator. These would be of interest in
designing a practical experiment, as it would be of consequence to
the number of counts that are obtained in the correlation
measurement. To this end, the fluorescence phonon emission spectrum
of the vdP oscillator is yielded by the power spectral density
(PSD), which is defined as the Fourier transform of the first-order
correlation function of the vdP oscillator field, namely,
$S(\omega)=\int_{-\infty}^{\infty}\langle\hat{a}^{\dag}(\tau)\hat{a}(0)\rangle
e^{-i(\omega-\omega_{d})\tau}d\tau$ \cite{V-5,V-6,V-7}, where the
angular brackets represent the average in the steady state of
Eq.\,(\ref{2}) and the calculation is performed in the frame
rotating with the driving frequency $\omega_{d}$. The PSD of the vdP
mode can be directly monitored in experiments.

In Fig.\,\ref{fig4}(a), we display the calculated phonon emission
spectra $S(\omega)$ as a function of the frequency $\omega$ for
seven different values of the two-phonon loss $\kappa_{2}$. As can
be seen, the phonon emission spectra possess a symmetrical
single-peak structure, which is of typical Lorentzian shape
reminiscent of a weakly driven two-level system \cite{V-5}. The peak
values are located at $\omega=\omega_{d}$, indicating
synchronization to the external driving. Here it is worth pointing
out that, without the nonlinear two-phonon loss and only with the
linear single-phonon loss corresponding to the case of
$\kappa_{2}/\kappa_{1}=0$ in Fig.\,\ref{fig4}(a), the spectral line
is flat and the phonon emission output is almost equal to zero. This
is due to the fact that the applied driving field is considerably
weak, $E_{d}=0.1\kappa_{1}$ in this linear oscillator system. When
$E_{d}$ is comparable to $\kappa_{1}$, the phonon emission output,
with a Lorentzian shape and a linewidth given by $\kappa_{1}$, can
occur (not shown). With increasing $\kappa_{2}$, the amount of
phonon emission is enhanced distinctly owing to the presence and
raise of the nonlinearity. Correspondingly, the spectral width of
the emission is increased. For gradually increasing $\kappa_2$ from
a zero value, the steady-state occupations of all Fock states
(including $|1\rangle$) decrease since there is an additional
relaxation channel opened, as is also confirmed by Eqs.\,(\ref{16})
and (\ref{17}) or Eqs.\,(\ref{27a}) and (\ref{27b}). Therefore, with
increasing the $\kappa_2/\kappa_1$ ratio we can expect an increasing
spectral width and emission amplitude.

In Fig.\,\ref{fig4}(b), we plot the amount of phonon emission
$S(\omega=\omega_{a})$ on the vdP oscillator resonance with the
driving (i.e., $\Delta=0$) as a function of $\kappa_{2}$ (left
scale). $S(\omega=\omega_{a})$ rises up quickly with $\kappa_{2}$.
For the sake of direct comparison with the above emission amount,
again we show the second-order correlation function $g^{(2)}(0)$ in
Fig.\,\ref{fig4}(b) (right scale). It is evident from these plots
that the changes of $S(\omega=\omega_{a})$ are opposite to
$g^{(2)}(0)$; strong phonon antibunching is accompanied by the high
amount of phonon emission, which is beneficial to the $g^{(2)}(0)$
measurement in practical experiments. For the sake of clarity, we
depict contour plot of $S(\omega)$ as a function of $\omega$ and
$\kappa_{2}$ in Fig.\,\ref{fig4}(c), the cuts of which through the
horizontal axis correspond to Fig.\,\ref{fig4}(a). In a word, the
stronger the antibunching is, the higher the collected fluorescence
(the power spectral density) is, as expected.

For all the discussions above, so far we only consider the
zero-time-delay second-order correlation function $g^{(2)}(0)$. Now
we focus on $g^{(2)}(\tau)$ with time delay $\tau$. Figure
\ref{fig5} depicts the behavior of $g^{(2)}(\tau)$ as a function of
both the time delay $\tau$ and the two-phonon loss $\kappa_{2}$ in a
2D contour plot. The inset in the main plot intuitively displays
$g^{(2)}(\tau)$ versus $\tau$ for various values of $\kappa_{2}$
shown in the legend. Note that, the delayed $g^{(2)}(\tau)$ is only
computed for $\tau\geq0$, because it is formally symmetric about
$\tau=0$, i.e., $g^{(2)}(-\tau)=g^{(2)}(\tau)$ as mentioned before.
We see from these figures that $g^{(2)}(\tau)$ monotonically
increases from a value smaller than $1$ (indicating antibunched
phonon emission) to $1$ (indicating coherent phonon emission) with
$\tau$ for a given nonzero value of $\kappa_{2}$. In particular, for
$\kappa_{2}=0$, $g^{(2)}(\tau)$ is identically equal to $1$. For a
sufficient $\tau$, $g^{(2)}(\tau)$ reaches the value of $1$. Most
obviously, a minimum value of $g^{(2)}(\tau)$ is obtained at
$\tau=0$. The minimum value of $g^{(2)}(\tau)$ diminishes with
increasing $\kappa_{2}$. Also, $g^{(2)}(\tau)$ persists at some
value below unity over a long time, on the order of $1/\kappa_{1}$.
From another point of view, it is found from Fig.\,\ref{fig5} that
the relationship $g^{(2)}(\tau)>g^{(2)}(0)$ is always satisfied for
$\kappa_{2}\neq0$, which violates the Cauchy-Schwarz inequality
$g^{(2)}(\tau)\leq g^{(2)}(0)$ related to classical light and thus
shows the phenomenon of phonon antibunching or sub-Poissonian
statistics related to nonclassical light \cite{V-5,V-6,V-7,V-10}.

\section{Feasibility of experimentally implementing a quantum vdP oscillator} \label{VI}
Regarding the experimental feasibility of our scheme, we note that
two typical setups, (i) an ion trap
\cite{VI-1,VI-2,VI-3,VI-4,VI-5,VI-6,VI-7} and (ii) an optomechanical
membrane \cite{VI-8, VI-9, VI-10}, have demonstrated the possibility
of implementing a vdP oscillator. First, for the ion trap setup
proposed originally in Refs.\,\cite{VI-1,VI-2}, the oscillator mode
$\hat{a}$ in Eq.\,(\ref{1}) stands for a linearly damped motional
degree of freedom of the trapped ion. Correspondingly, the linear
one-phonon damping can occur when the standard laser cooling
techniques are utilized \cite{VI-8}. One can engineer the nonlinear
damping via applying sideband transitions to remove energy quanta,
for example, the two-phonon loss can be carried out by laser
exciting a harmonically trapped ion to its red motional sideband by
removing two phonons at a time \cite{VI-1,VI-2}. In this scenario,
the one- and two-phonon loss rates are both of the order of kHz,
with $\kappa_{2}\geq\kappa_{1}$ \cite{VI-1,VI-2,VI-3,VI-6,VI-7}. As
also shown in Refs.\,\cite{VI-1,VI-7}, it is possible to resolve
sidebands and suppress off-resonant excitations for several tens of
low-energy modes.

Second, for the optomechanical ``membrane-in-the-middle'' system
\cite{VI-9,VI-10,VI-11} or the mechanical self-oscillation in cavity
optomechanical system \cite{VI-12,VI-13,VI-14}, one can take into
account a moderate quality factor membrane with the linear
mechanical damping. The nonlinear two-phonon damping can be
implemented by exploiting a laser red-detuned with two mechanical
frequencies relative to the two-phonon sideband. One can apply an
electric field gradient created near the membrane to create the
driving force.

As a side note, the present phonon mode can be extended to a
stationary photonic mode in an optical system, such as a microwave
resonator or an optical cavity. Strong two-photon loss can also be
realized successfully through the Josephson junction in
superconducting circuit with high tunability \cite{VI-15} or through
using an on-chip toroidal microcavity immersed in rubidium vapor
\cite{I-74}.

Finally, one can use the objective to collect the phonon or photon
emission from the oscillator mode $\hat{a}$. Then it is sent to a
Hanbury-Brown--Twiss (HBT) configuration, consisting of a $50:50$
beam splitter and a pair of single-phonon or single-photon counting
detectors. The normalized second-order correlation function of
interest, characterizing the quantum statistics of particle sources,
can be measured by this HBT setup (see, for instance,
Refs.\,\cite{V-5,V-6,V-7}, and references therein). Thus, we expect
that an experimental implementation of our scheme is feasible in the
present state of the art.

\section{Conclusions} \label{VII}
In summary, we have introduced an alternative route towards highly
nonclassical phonon emission statistics in a driven quantum vdP
oscillator subject to both the linear one-phonon and nonlinear
two-phonon damping by evaluating the second-order correlation
function $g^{(2)}$ numerically and analytically. With experimentally
achievable parameters, strong antibunching in the emitted phonon
statistics is revealed by the full numerical simulations using a
quantum master equation, which is in good agreement with analytical
calculations utilizing a three--vdP-oscillator--level model by
assuming that only the three lowest Fock states of the vdP
oscillator have non-negligible occupations under weak driving. It is
shown that the degree of phonon antibunching is limited by the rate
of the two-phonon loss in this vdP oscillator, and the antibunching
is enhanced considerably with growing the two-phonon loss. So, the
two-phonon loss is very important for the realization of a good
single-phonon device. The degree of phonon antibunching can be kept
high even when $\kappa_{2}<\kappa_{1}$. Suitable conditions for
strong phonon antibunching generation and single-phonon emission are
identified in the parameter regimes. On the other hand, we also
numerically calculate the fluorescence phonon emission spectra
$S(\omega)$, given by the power spectral density of the driven vdP
oscillator. We further find that high phonon emission amplitudes,
simultaneously accompanied by strong phonon antibunching, can be
obtained, which are beneficial to the HBT measurement of the emitted
phonons in practical experiments. Lastly, following the original
proposals in Refs.\,\cite{VI-1,VI-2}, we illustrate specific
implementations applying a harmonically trapped ion and an
optomechanical membrane in the middle. Our model's description of
the interplay between one-phonon and two-phonon losses and their
effects is not limited to phononic systems and should be generally
applicable to photonic systems. This vdP architecture provides a
feasible way of realizing efficient single-phonon or single-photon
sources for quantum information processing tasks.

\section*{Acknowledgements}
The help of the two anonymous referees in improving this paper is
gratefully acknowledged. The authors thank Rong Yu for fruitful
discussions during the paper preparation. J.L. was supported in part
by the National Key Research and Development Program of China under
Contract No.\,\,2016YFA0301200, by the National Natural Science
Foundation of China through Grant No.\,\,11675058, and by the
Fundamental Research Funds for the Central Universities (Huazhong
University of Science and Technology) under Project
No.\,\,2018KFYYXJJ037. C.D. was supported by the National Natural
Science Foundation of China through Grants No.\,\,11705131 and
No.\,\,U1504111, as well as by the Science Research Funds of Wuhan
Institute of Technology under Project No. K201744. Y.W. was
supported partially by the National Natural Science Foundation of
China through Grants No.\,\,11875029 and No.\,\,11574104.

\section*{Data Availability}
The data that support the findings of this study are available from
the corresponding author upon reasonable request.

\end{document}